\documentclass[12pt]{article} \usepackage{amsfonts}
\usepackage[dvipsnames]{xcolor}
 \usepackage{amsmath,amssymb,amsthm,epsfig} 

\input colordvi.sty
\usepackage{xcolor}
\usepackage{graphicx}
\usepackage{datetime}
\usepackage{cleveref}
\usepackage{pdfsync} 
\DeclareSymbolFontAlphabet{\mathbb}{AMSb}
\begin{document}
\renewcommand{\theequation}{\thesection.\arabic{equation}}
\newcommand{\R}   {\mathbb{R}} \newcommand{\N}   {\mathbb{N}}
\newcommand{\C}   {\mathbb{C}} \newcommand{\Z}   {\mathbb{Z}}
\newcommand{\0}   {\mbox{\bf 0}} \newcommand{\alert}
{\marginpar{\Large ???}} \newcommand{\res}    {\mbox{\rm res}}
\newcommand{\eproof} {\framebox{}} \newcommand{\app
}[1]{\stackrel{\mbox{\scriptsize \rm(A.{#1})}}{=}} \newcommand{\V}
{\mbox{\it V}} \newcommand{\U}     {\mbox{\it U}} \newcommand{\vv}
{\mbox{\it v}} \newcommand{\uu}     {\mbox{\it u}}
\newcommand{\ww}     {\mbox{\it w}} \newcommand{\Skw}
{\mbox{\rm Skw}\,} \newcommand{\Sym}    {\mbox{\rm Sym}\,}
\newcommand{\skw}     {\mbox{\rm skw}\,} \newcommand{\sym}
{\mbox{\rm sym}\,} \newcommand{\grad}    {\mbox{\rm grad}\,}
\newcommand{\divt}    {\mbox{\rm Div}\,} \newcommand{\divv}
{\mbox{\rm div}\,}
\newcommand{\Dym}    {\mbox{\scriptsize \rm Dym}}
\newcommand{\ARTICLE}[1]    { \bibitem{#1}} \newcommand{\BOOK}[1]
{ \bibitem{#1}} \newcommand{\AUTHOR}[1]{{#1}:}
\newcommand{\JOURNAL}[1]{{#1},} \newcommand{\TITLE}[1]{{\it{#1}},}
\newcommand{\VOLUME}[1]{{#1},} \newcommand{\PAGES}[1]{p. {#1},}
\newcommand{\PUBLISHER}[1]{{#1},} \newcommand{\ADDRESS}[1]{{#1},}
\newcommand{\YEAR}[1]{{#1}.}

\newcounter{mycounter}[section]
\def\themycounter{\thesection.\arabic{mycounter}}
\newcommand{\definition}[1]{\refstepcounter{mycounter}
\vspace{3.0ex}
 \noindent{\bf Definition \themycounter~~} {\sl #1}\par   \vspace{3.0ex}}
\newcommand{\lemma}[1]{
\vspace{3.0ex}
 \noindent{\bf Lemma \themycounter~~} {\sl #1}\par     \vspace{3.0ex}}
\newcommand{\theorem}[1]{\refstepcounter{mycounter}
\vspace{3.0ex}
 \noindent{\bf Theorem \themycounter~~} {\sl #1}\par   \vspace{3.0ex}}
\newcommand{\remark}[1]{\refstepcounter{mycounter}
\vspace{3.0ex}
 \noindent{\bf Remark }
{\rm #1 {~}\hfill \eproof\par\goodbreak\vspace{3.0ex}}}
\newcommand{\proposition}[1]{\refstepcounter{mycounter}
\vspace{3.0ex}
 \noindent{\bf Proposition \themycounter~~} {\sl #1}\par    \vspace{3.0ex}}
\newcommand{\example}[1]{\refstepcounter{mycounter}
\vspace{3.0ex}
 \noindent{\bf Example \themycounter:~} {#1}\par   \vspace{3.0ex}}
\newcommand{\Problem}[1]{\refstepcounter{mycounter}
\vspace{3.0ex}
 \noindent{\bf Problem \themycounter :~} {#1}\par   \vspace{3.0ex}}

\bigskip\bigskip\bigskip \begin{center}
{\Large \bf Some remarks on 
the model of  rigid heat conductor with memory: unbounded heat relaxation 
function}
\end{center}
\normalsize \vspace{0.2cm}

\medskip
\begin{center}

{\bf Sandra Carillo
\\{\rm Dipartimento Scienze di Base e Applicate per l'Ingegneria, 
\\
      Universit\`a di Roma ``La Sapienza'',
      I-00161 Rome, Italy \\\ \& \\
                        { ~I.N.F.N. - Sezione Roma1, Gr. IV}\\ {Mathematical Methods in NonLinear Physics,  Rome, Italy}}
\\  \vspace{0.2cm} {\rm  March, 23 2018}
} \end{center}
\medskip \begin{abstract} \noindent The model of rigid linear heat conductor
with memory is reconsidered focussing the  interest on the 
heat relaxation function. 
Thus, the  definitions of heat flux and thermal work 
are revised to understand where changes are required when the heat flux relaxation 
function $k$ is assumed to be unbounded at the initial time $t=0$. That is, it is represented by
a regular integrable function, namely $k\in L^1(\R^+)$, 
but its time derivative is not integrable, that is $\dot k\notin L^1(\R^+)$.
Notably, also under these relaxed assumptions on $k$, whenever the heat flux is the
same also the related thermal work is the same. Thus, also in the case under investigation,
the notion of equivalence is introduced and its physical relevance is pointed out. 
\end{abstract} \vspace{2.0ex}

{\small \noindent MCS numbers: {\bf 80A20,
74F05}
\\
\vspace{2.0ex} 

\smallskip \section{Introduction} \setcounter{equation}{0} 

The model of a rigid heat conductor is well known: its origins can be traced 
in the celebrated generalised Fourier's law by Cattaneo \cite{Cattaneo}.
The approach here adopted is based on subsequent results by Coleman
\cite{Coleman}, concerning materials with memory, which were further developed 
by Gurtin and Pipkin \cite{GurPip}, who  proposed
a non-linear generalisation  in the case of a rigid heat conductor with memory.
Since then, heat transfer phenomena have been widely studied, to mention
only some of the results connected to the present study,  
 Gurtin \cite{Gur} and Coleman and Dill \cite{ColemanDill}  studied
 properties of free energy functionals in the case of materials with memory. 
Later, Giorgi and Gentili \cite{GioGen} further investigated, in a fading memory
material, a problem in heat conduction. The model under investigation is 
that one of a rigid homogeneous linear heat conductor with memory studied by
Fabrizio, Gentili and Reynolds  \cite{FGR} considered in \cite{SAM}. 
Specifically, the model in \cite{SAM} is revisited to introduce those generalisations which are 
required to adapt the model itself to describe a wider class of materials with memory. 
To start with the assumptions on the physical quantities of interest to describe the problem are briefly summarised. 
 The  internal energy and the relative temperature are 
assumed linearly related, that is, 
\begin{equation}\label{1}
e ({\bf x}, t) = \alpha_0  u ({\bf x}, t)~~,
\end{equation} 
where, respectively, $e$ denotes  the  internal energy,
$\alpha_0$  the {\it energy relaxation function},  for  simplicity, 
assumed to be constant, ${\bf x} \in{\cal B}\subset\R^3$ where ${\cal B}$ denotes
the bounded closed set in $\R^3$ which represents
the configuration domain of the conductor, denotes
the position within the conductor, 
$t\in\R^+$ denotes the time
variable\footnote{{\bf notation remark:}
throughout the whole paper $\R^+:= [0, \infty)$
while $\R^{++}:= (0, \infty)$.}, and $ u :=\theta
- \theta_0 $ the temperature difference with
respect to a fixed reference temperature
$\theta_0 $.
According to the {\it classical} setting, see
Giorgi and Gentili 
\cite{GioGen} and also Fabrizio, Gentili and Reynolds \cite{FGR},  the heat flux ${\bf q} \in \R^3$ is assumed to satisfy
the constitutive equation
 \begin{equation}\label{2} {\bf q} ({\bf x}, t) =
-\displaystyle{\int_{0}^{\infty} { k(\tau) {\bf g}({\bf x}, t-\tau) ~d \tau }}~~,
\end{equation} where ${\bf g}:=\nabla {u}= \nabla (\theta
- \theta_0) =\nabla \theta $ is the
temperature-gradient and $k(\tau)$ the heat flux
relaxation function, which is assumed
 \begin{equation}\label{k-eq} k(t) = k_0 +
\displaystyle{\int_{0}^{t} { \dot k(s) ~d s }}~~, \end{equation}
where $k_0\equiv k(0)$ represents the initial (positive) value
of the heat flux relaxation function, thus termed {\it initial
heat flux relaxation coefficient}. Under the functional viewpoint, 
the heat flux relaxation function is classically assumed to satisfy the 
following regularity requirements:
\begin{equation}\label{3b} \dot k \in L^1 (\R^+) \cap  L^2  (\R^+)
\end{equation}
and 
\begin{equation}\label{k}
k \in L^1 (\R^+)~.
\end{equation} 
These requirements are consistent with the physically meaningful condition of no heat flux when, 
at infinity, the thermal equilibrium is reached. Hence, the asymptotic behaviour of  the heat flux relaxation function, at $t\to \infty$, is assumed 
$\displaystyle{k(\infty):=\lim_{t\to \infty} k(t)=0}$.

\noindent
On introduction of the {\it integrated history of the temperature-gradient}
\begin{equation}
\bar{\bf g}^t(\tau)=
\displaystyle{\int_{t-\tau}^{t} {\bf g}(s) ~d s}\end{equation}
 the heat flux \eqref{2.6}, can be written also as
\begin{equation}\label{2b}
\displaystyle{
{\bf q}(t) =
\int_{0}^{\infty} { \dot k(\tau)
\bar{\bf g}^t(\tau) ~d \tau }~~.}
\end{equation}
Then, the evolution equation which describes the temperature evolution within a rigid heat 
conductor with memory,  reads
\begin{equation}\label{eq00}
u_t= -\nabla \cdot {\bf q}({\bf x}, t) + r({\bf x},t)~,
\end{equation}
wherein  the heat flux $ {\bf q}$ is given by \eqref{2} or, equivalently, by \eqref{2b}, hence, 
in case of no heat supply, in turn
\begin{equation}\label{eq00}
u_t= \nabla \cdot \int_{0}^{\infty} { k(\tau) {\bf g}({\bf x}, t-\tau) ~d \tau } ~~\textstyle{or}~~
u_t= \nabla \cdot \int_{0}^{\infty} { \dot k(\tau)
\bar{\bf g}^t(\tau) ~d \tau }.
\end{equation}
Note that in the two linear integro-differential equations, respectively, the kernel coincides with the 
heat flux relaxation function or its time derivative: both of them, in the classical case, i.e. when $k$ 
satisfies both \eqref{k} and \eqref{3b}, are regular.  This condition is not satisfied in the present 
investigation since the heat flux relaxation function $k$ is assumed to be unbounded at $t=0$.
Nevertheless, the solution existence problem can be addressed to: some results are comprised in 
\cite{CVV2013a, MECC2015,  [95]}, now the interest is focussed on the model and, specifically to the definition of heat flux and thermal work, in this generalised case. 

The existence  and uniqueness result of solution to \eqref{eq00} when Dirchlet boundary conditions 
and  assigned initial data takes its origin in the results   
by Dafermos \cite{Dafermos, Dafermos1}  in connection  
with an analogous problem which arises in linear viscoelasticity. Indeed, under the analytical viewpoint, 
the two different physical models of isothermal viscoelasticity and rigid heat conduction with memory 
share interesting properties as pointed out, for instance, in \cite{S2005,JNMP2015}. 
The interest in singular kernel models is not new since the seminal idea is due to  Boltzmann \cite{Boltzmann} in connection with 
special viscoelastic behaviours, further references  are comprised in \cite{Intech2016}. 
The term   {\it singular kernel} problem is introduced to refer to this  model to stress that  \eqref{eq00}
 is  characterised by a  kernel unbounded at $t=0$. The interest in {\it singular kernel} problems is well 
 known both under the analytical viewpoint \cite{Grasselli, Janno1, Janno2, Miller-Feldstein, CVV2013a} 
 as well as in connection to the study of innovative materials \cite{CPV} or materials whose response is 
 changed due to {\it aging} processes \cite{Intech2016}  or models devised to describe bio-materials  
 \cite{Deseri}. As a special case, the current interest is on models in which the singular kernel is 
 represented by a fractional derivative term   \cite{Fabrizio2014, Mainardi, de Andrade}.

\medskip
The article is organised as follows.
The opening Section 2 is concerned about the notion of process and process prolongation. 
Specifically, the definitions given in \cite{SAM} are revised and adapted to take into account 
that the integrability of the time derivative of the  heat  flux relaxation function, assumed in 
\cite{SAM}, is not required in the present study. 

The following Section 3 is devoted to revise the expression of the  heat flux to adapt it to the  
{\it singular} case under investigation.
Again, the required changes are introduced. Furthermore, in Section 3, the notion of equivalence 
among different thermal 
processes is introduced. It represents a generalisation of the analogous one which refers to the
 regular model \cite{SAM} 
and allows to identify thermal processes which shares the same heat flux.  

The thermal work is considered in Section 4 where the notion of equivalence is given in terms 
of the thermal work. Notably, in the singular as well as in the classical model, the crucial property 
that  processes which are associated 
to the same heat flux are also associated to the same thermal work holds true. Hence, equivalence 
can be stated, indifferently, referring to one or the other physically meaningful quantities.
Conclusions and perspective investigations are comprised in the closing Section 5. \medskip

\section{Process \& process prolongation}
The aim of this section is  to revise the notions introduced in \cite{SAM}
to adapt them to the case of a singular kernel material. 

Accordingly, the definition of all the meaningful quantities is provided 
with the exception of  those ones which are exactly the same in the regular as
well as in the singular cases.

The present investigation, as \cite{SAM} is based on the model by Fabrizio,
Gentili and Reynolds in \cite{FGR}, hence the attention is
focused on an element of the conductor which is supposed  isotropic. 
No dependence on the position in the conductor is considered so that 
the energy $e$, the heat flux ${\bf q}$, the
temperature $\theta$ and also all the other derived
quantities  are represented by  functions of the time variable
alone. Conversely, all such quantities depend on the time variable  not only via the {\it present} 
time $t$, but also via their past {\it history}.

This Section collects some crucial definitions, according to the model in  \cite{FGR}, 
such as  process prolongation, equivalent histories and the {\it zero prolongation} 
 recalled from  \cite{SAM}  since they represent key notions throughout.
Some slight modifications are needed, which, however, do not change the results.
Specifically,  integrals wherein $\dot k$ appears, which loose their meaning when 
$\dot k \notin L^1$,  are avoided. The presence of these minor changes represents a
further justification for  this  Section whose aim is exactly to point out those definitions 
which need to be modified (and how)  with respect to the analogous ones comprised 
in \cite{SAM}.

 {\definition{

\noindent An integrable function $P:[0,T) \to \R
\times \R^3$ such that $P(\tau)= \left( \dot
\theta_P(\tau), {\bf g}_P(\tau) \right)$ $
\forall~\tau \in [0, T)$ is termed {\it process
of duration} $T>0$.}}

\noindent Hence, a process $P$ of duration $T<\infty$ is known
when the two applications $\dot\theta_P :[0,T) \to \R$ and ${\bf
g}_P~:[0,T) \to \R^3$ are assigned, and thus the state function
$\sigma(t)$ can be constructed.  
Indeed, as pointed out in \cite{FGR}, given its initial value
$\sigma(0) =\left( \theta_{\star}(0),  
\bar{\bf g}_{\star}^0\right)$, where, in turn,
$\theta_{\star}(0)$ denotes the temperature,
and $\bar{\bf g}_{\star}^0$ the integrated
history of the temperature-gradient at time
$t=0$, then a process $P$ delivers the
state function $\sigma(t)$,$ \forall~t\in
[0,T)$, defined by
\begin{equation}\label{nota1} \displaystyle{ \theta(t)=
\theta_{\star}(0)+ \int_0^t \dot \theta_P(\xi)~d\xi ~~,}
\end{equation}
 and
\begin{equation}\label{nota3} \bar{\bf g}^t(s)=\left \{
\begin{array}{ll} \displaystyle{\int_{t-s}^{t} {{\bf g}_P (\xi)
~d
\xi }}  &
0\le s < t\\[2.0ex] \displaystyle{\int_{0}^{t} {{\bf g}_P(\xi)
~d
\xi + \bar{\bf g}_{\star}^0(s - t) }} &s \ge t~,
\end{array}\right . \end{equation}
which is continuous at $s=t$.
To specify what it means to define a
``prolongation'' of a generic process, whose history is known,
with a given one, characterized by an assigned ${\bf
g}_P$, the following definition is introduced.

{\definition{

\noindent Given a process $P$, hence assigned
${\bf g}_P$, its duration $T$ and a set of
integrated histories of the temperature-gradient
$\bar{\bf g}_i^t(s)$, corresponding to ${\bf
g}_i^t(s), ~i=1,\dots,n,~ n\in \N$, then the
prolongation of the history $\bar{\bf g}_i^t(s)$,
induced by the process $P$, is defined by
\begin{equation}\label{prolong}
\!\left( {\bf g}_P \star
\bar{\bf g}_i\right)^{t+T}(s): =\!\!\left \{ \!\!\begin{array}{ll}
\displaystyle{ \bar{\bf g}_P^{T}(s)=\!\int_{T-s}^{T} {{\bf
g}_P (\xi) ~d
\xi}}  & 0\le s < T\\[2.0ex] \displaystyle{
\bar{\bf g}_P^{T}(T) +
 \bar{\bf g}_{i}^t(s - T) =\!
\int_{0}^{T}\!\!\!\!\!\!{{\bf g}_P(\xi) ~d
\xi + \int_{t+T-s}^{t} \!\!\!\!\!\!{{\bf g}_i(\xi) ~d
\xi }
}} &~~~~ s \ge T~,
\end{array}\right.
 \end{equation} }}
 \noindent
 which is continuous at $s=T$ since  $\left( {\bf g}_P \star
\bar{\bf g}_i \right)^{T}(T)=\bar{\bf g}_P^{T}(T)$ and 
$\left( {\bf g}_P \star
{\bf g}_i \right)^{T}(T)={\bf g}_P^{T}(T)$.
Such a definition can be  equivalenly 
introduced in terms of the translated temperature 
gradient instead than of the integrated history of the temperature 
gradient. Thus, the  definition of prolongation of the  translated temperature 
gradient reads as follows.

{\definition{

\noindent Given a process $P$, hence assigned
${\bf g}_P$, its duration $T$ and a set of
translated  temperature-gradient
${\bf g}_i^t(s)$, corresponding to ${\bf g}_i^t(s), ~i=1,\dots,n,~ n\in \N$, 
then the prolongation of  ${\bf g}_i^t(s)$,
induced by the process $P$, is defined by
\begin{equation}\label{prolong2}
\!\left( {\bf g}_P \star
{\bf g}_i\right)^{t+T}(s): =\!\!\left \{ \!\!\begin{array}{ll}
\displaystyle{ {\bf g}_P^{T}(s)= {\bf g}_P ({T}-s)={\bf g}_P ^{T}(s)}
& 0\le s < T\\[2.0ex] \displaystyle{
{\bf g}_i (t+T-s)={\bf g}_i^{t+T}(s) 
} &~~~~ s \ge T~,
\end{array}\right.
 \end{equation}
 }}

\noindent 
which is continuous letting ${\bf g}^t_i(0)={\bf g}^T_P(0)$.
As a special case, ${\bf g}_{0 }^t={\bf 0}$ termed {\it zero history} in \cite{SAM},  
characterised by a constant  temperature $\theta(t)=\theta_0$ and  a zero 
temperature-gradient  is introduced: the corresponding heat flux is zero.
The corresponding state function is given by
 \begin{equation}\label{zerohistory}
\begin{array}{cc}
 ~~~~~~
\sigma_0(t):&\!\!\!\!\!\!\!\!\!\!\!\!\!
\!\!\!\!\!\!\!\!\!\!\!\!\!\!\!\R\to
\R
\times 
\R^3 \\
    & t \longmapsto \sigma_0(t) \equiv\left(
\theta_0, {\bf 0}\right)~.\\
 \end{array}
\end{equation}
Consider, now, a prolongation of the zero history via any
assigned process, characterized by the duration $\tau$, $\tau \le T< \infty$
and  ${\bf g}_P: [0, \tau ) \to \R^3$, then
\begin{equation}\label{zero_prol}
 \left( {\bf g}_P \star \bar{\bf g}_0
\right)^{t+ \tau }(s) =\left \{
\begin{array}{ll} \displaystyle{ \bar{\bf g}_P^{
\tau }(s)=\int_{ \tau -s}^{ \tau } {{\bf g}_P
(\xi) ~d
\xi }} ~~~~ & 0\le s<\tau \\[2.0ex] \displaystyle{\bar{\bf g}_P^{ \tau}(\tau) } &~~~~ s \ge\tau ~.
\end{array}\right.
 \end{equation}

\section{Heat flux \& equivalence}
\label{section3} \setcounter{equation}{0}
The aim of this section is  to revise the notions introduced in \cite{SAM}
to adapt them to the case of a singular kernel material. 

 This Section is  devoted to the heat flux  functional and its expression in the
 case the heat flux relaxation function $k$ satisfies \eqref{k}, but does not 
 satisfy \eqref{3b}.
 The constitutive equations
({\ref{1}}) and ({\ref{2}}) are considered. The latter, which represents 
the heat flux, on introduction of  ${\bf g}^t(\tau):= {\bf g}(t-\tau) $
termed {\it translated temperature-gradient}, can be written as
\begin{equation}\label{2.6}
{\bf q} (t) =
-\displaystyle{\int_{0}^{\infty} { k(\tau) {\bf g}(t-\tau)
~d \tau}  ~~~ \text{i.e.} ~~~ {\bf q} (t) = - \int_{0}^{\infty} { k(\tau) {\bf g}^t(\tau)
~d \tau ~~}}~~.
\end{equation} 
Such a definition  implies that the heat flux at the  time $t+T$, for
    any $T>0$, reads
    \begin{equation}\label{81} \displaystyle{  {\bf q}(t+T):= -
    \int_{0}^{\infty} {  k(s) \,{\bf g}^{t+T}(s) ~d s}~. }
    \end{equation} 

Nevertheless,  following the spirit of Fabrizio, Gentili and
Reynolds \cite{FGR} and McCarthy \cite{McCarthy}, a  {\it thermodynamic state function} can be introduced also in the singular case.
Hence, as a natural modification of definition 2.1 in \cite{SAM} (see also \cite{FGR}),  the thermodynamical state of the conductor is assigned  via the following {\it thermodynamic state function}.
\footnote{Note that the adoption of this definition or of definition 2.1 in \cite{SAM} 
are completely equivalent in the regular case when $\dot k\in L^1(\R^+)$.}

\medskip
{\definition{The function 
 \begin{equation}\label{history}
\begin{array}{cc}
 ~~~~~~
\sigma(t):&\!\!\!\!\!\!\!\!\!\!\!\!\!
\!\!\!\!\!\!\!\!\!\!\!\!\!\!\!\R\to
\R
\times 
\R^3 \\
    & t \longmapsto \sigma(t) \equiv\left(
\theta(t), {\bf g}^t\right)~\\
 \end{array}
\end{equation}
where ${\bf g}^t$ belongs to a suitable Hilbert space, is said to be the
{\it thermodynamic state function}: it characterises the
thermodynamic state  of the conductor.}}

\medskip
\noindent
Therefore, the thermodynamic state
function is known as soon as the temperature and the translated
 temperature-gradient are given. The heat flux linear functional can be defined via: 
\begin{equation}\label{8c}
\displaystyle{\widetilde Q \{{\bf g}^t\}:= -\int_{0}^{\infty} {
 k(s) \,{\bf g}^t(s) ~d s}~~ \Longrightarrow~~ \displaystyle{ \widetilde Q \{{\bf g}^{t+T}\}= -
    \int_{0}^{\infty} {  k(s) \,{\bf g}^{t+T}(s) ~d s}~, ~~\forall T>0. }}
\end{equation} 

\noindent
Hence, since only finite heat fluxes  are physically admissible, the set of all  ${\bf g}^t$ 
which correspond to  them belong to the vector space\footnote{This definition is also 
slightly different with respect to the corresponding one (see formula  (2.13) in 
\cite{SAM}), however the  two conditions which characterise the space  $\Gamma$ do 
coincide in the regular case. The different consists in the elements of the space  $
\Gamma$ since, in the singular case, the heat flux is more conveniently expressed in 
terms of the translated temperature-gradient  ${\bf g}^t$ instead than of the integrated history of the temperature-gradient  $\bar{\bf g}^t$ adopted in \cite{SAM}.}
\begin{equation}\label{8b} \displaystyle{\Gamma:= \left\{
{\bf g}^t : (0, \infty) \to \R^3 :~
\left\vert~\int_{0}^{\infty} {  k(s+\tau)\, {\bf g}^t(s)
~d s} ~\right\vert < \infty~~,~~\forall ~\tau \ge 0\right\}}~.
\end{equation}
 Correspondingly,  the set of states associated  to a finite heat flux is 
\begin{equation}\label{Sigma}
\displaystyle{\Sigma:= \left\{ \sigma(t) \in \R \times \R^3\,:
\left\vert~\int_{0}^{\infty}\!\!\!\! { k(s) \,{\bf g}^t(s) d s} \right\vert <
\infty\right\}}~. 
\end{equation} 
{\bf {Remark}}  The {\it fading memory} property holds true also in the singular case, 
however, when $\dot k \notin L^1(\R^+)$, the condition to impose to
guarantee  that,   consistently with the prescription there is no heat flux when the
asymptotic equilibrium is reached,  follows if, given any arbitrary $\varepsilon >0$,
there exists a positive constant $\tilde a=a(\varepsilon, {\bf g}^t)$ s.t.
\begin{equation}\label{9}
\displaystyle{\left\vert~\int_{0}^{\infty} {  k(s+a) {\bf g}(t-s)
~d s} ~\right\vert < \varepsilon~~,~~\forall\, a > \tilde a}~~~,
\end{equation}
which coincides with the usual condition, see (2.14) in \cite{SAM},
in the regular case when $k$ satisfies also \eqref{3b}. %

Hence, the notion of equivalence can be given to identify all those
 translated temperature gradient histories ${\bf g}^t$ which correspond to the
 same heat flux.

{\definition{\label{eq2}

\noindent Given two translated temperature 
gradient histories ${\bf g}^t_1$ and ${\bf g}^t_2$  are said to be {\it equivalent} if
\begin{equation}\label{16} \displaystyle{
\forall~ {\bf g}_P~: [0, \tau) \to
\R^3~~,~~\forall~~\tau\in[0, T) }
\end{equation} they satisfy
\begin{equation}\label{13c}
\displaystyle{ \widetilde Q \left\{ ({\bf g}_P
\star {\bf g}_1)^{t+\tau}\right \}= \widetilde Q
\left\{ ({\bf g}_P \star {\bf g}_2)^{t+\tau}
\right\}~~. } \end{equation} }}

That is, two different  histories of the
temperature-gradient, or two thermodynamical
states, are equivalent when, for any prolongation
of any time duration, the same heat flux
corresponds to both of them. 
The physical meaning, which naturally does not dependent on the regularity 
of the heat flux relaxation function, is to recognise those thermal states which
correspond 
to the same heat flux, exactly as stated in \cite{SAM}. 
Hence, according to \eqref{13c}, recalling \eqref{prolong2}, equivalent states produce 
the same heat flux as soon as, for any $s \ge  \tau$, it holds 
\begin{equation}\label{14}
\displaystyle{ \int_{0}^{\infty}\!\!\!\!  k(s)
 \left( {\bf g}_P \star {\bf
g}_2\right)^{t+ \tau }(s) ds =
\int_{0}^{\infty}\!\!\!\!  k(s)
\left( {\bf g}_P \star {\bf g}_1\right)^{t+ \tau }( s)
ds }~,
\end{equation}
which, on substitution of (\ref{prolong2}), reads  
\begin{equation}\label{14b}
\displaystyle{
  \int_{ \tau }^{\infty}\!\!\!\!  k(s)  {\bf g}_{2
}^{t}(s-\tau) ds =
\int_{\tau}^{\infty}\!\!\!\!  k(s){\bf g}_{1}^{t}(s-\tau)
 ds}~.
\end{equation} The latter shows also that two translated histories of the
temperature-gradient ${\bf g}^t_1$ and ${\bf g}^t_2$  are equivalent whenever 
their difference ${\bf g}^t_1- {\bf g}^t_2$ is equivalent to zero, that is 
\begin{equation}\label{zero_cond}
\displaystyle{ \int_{ \tau }^{\infty}  k(\xi) ~ {\bf g}^{t}( \xi-\tau) ~ d\xi= 0 ~,~~ {\bf g}^t:= {\bf g}^t_1-{\bf g}^t_2}~.
\end{equation}
 Furthermore, when  the couples
 $(\theta_1(t), {\bf g}_1^t(s)) $ and
 $(\theta_2(t), {\bf g}_2^t(s)) $
 are considered then,
they can be represented by the same state
function  $\sigma(t)$ whenever they are such that
 $\theta_1(t)=\theta_2(t)~~\forall~ t$ and,    in addition,
 the difference of the two  integrated histories of the
temperature-gradient is equivalent to zero, i.e.
satisfies (\ref{zero_cond}).

\section{Thermal work}
In this Section  the notion of  {\it thermal work}, in the singular case,  is introduced 
in such a way that it coincides with the corresponding one  in the regular case \cite{SAM}.
To introduce the  definition of  thermal work, according to \cite{FGR}, 
the notions of {\it process} and process {\it  prolongation}, given in Section 2, are  
needed.  A comparison between the definitions in Section 2 and the corresponding ones in 
\cite{SAM} (Definitions 2.2 and 2.3) shows that some differences turn out to appear. Indeed, 
even if both the definitions introduced  in \cite{SAM} do not depend on the heat flux relaxation 
function, and, hence, can be adopted also in the singular case, this choice is not convenient.
The new definitions in Section 2 are suggested by the need to use, to express the heat flux as
well as the thermal work,  the heat flux relaxation function instead than its time derivative; thus,
the key role played in  \cite{SAM} by the integrated history of the temperature-gradient 
$\bar{\bf g}^t$ is now of the translated temperature  gradient ${\bf g}^t$.
 
As fas as the  thermal work is concerned,  the difference between the two cases regular and singular 
 is limited to the fact that the expression of the thermal work can be given in terms of
 the heat flux relaxation function while expressions in which contain integrals of its time 
 derivative are not defined.

{\definition{
 {
\noindent Given any initial state,
defined on assigning the  state function 
$\sigma(t)=(\theta(t), \bar{\bf g}^t)$, and any prolongation
process $P$ of arbitrary finite duration $\tau$,
  $P(\tau)= \left( \dot
  \theta_P(\tau), {\bf g}_P(\tau) \right)$, where $ 0\le \tau< T$,
then the thermal work associated to the time
interval $[0,T)$ is represented by the
functional 
\begin{equation}\label{23}
\displaystyle{\widetilde W \{  {\bf g}^t;
 {\bf g}_P\}:= -\int_{0}^{T} {  \widetilde Q \left\{ ({\bf g}_P \star
{\bf g})^{t+\tau}\right \}
 \cdot {\bf g}_P(\tau) ~d \tau }}~~.
\end{equation} }}
\noindent
Thus, chosen ${\bf g}_P $ and the initial state,  the thermal work  
$\widetilde W ( {\bf g}^t; {\bf g}_P)$  follows
\begin{equation}\label{23z}
\displaystyle{\widetilde W \left( {\bf g}^t;
 {\bf g}_P\right)= -\int_{0}^{T} { {\bf q} \left({t+\tau}\right)
 \cdot {\bf g}_P(\tau) ~d \tau }}~~
\end{equation}
recalling the expression (\ref{81}) of the heat flux ${\bf q} \left({t+\tau}\right)$ at time
$t+\tau$. If, in particular,   the prolongation, via any assigned process
$P$, of  the {\it zero history}, which corresponds to the state function
$\sigma(t)\equiv\left( \theta_0,{\bf 0}\right)$, is considered, then
the thermal work functional, according to \cite{SAM}, is 
 
 \begin{equation}\label{230}
 \displaystyle{\widetilde W \{ {\bf 0};
{\bf g}_P\} =
\int_{0}^{\infty}\!\!\! \int_{0}^{\tau}  {
 k(s) {\bf g}_P({\tau} - s)} \cdot {\bf g}_P(\tau)
 \, d s    d \tau}~~.\end{equation} 

Notably, the expression (\ref{230}) of the thermal work
$\widetilde W\{ {\bf 0}; {\bf g}_P\}$, remains valid
for any given process $P$, that is any ${\bf
g}_P(\tau)$, and $\forall ~0\le \tau < T< \infty$.
The  thermal work $\widetilde W\{ {\bf 0};
{\bf g}_P\}$ on introduction, in (\ref{230}), of  $\xi= \tau -s$,  and, then,
$ s= \xi$, can also be written as
\begin{equation}\label{3.3}
\displaystyle{ \widetilde W \{ {\bf 0};
{\bf g}_P\}=
\int_{0}^{\infty}\!\!\!\! \int_{0}^{\tau}\!\!  {
 k({\tau} - s) {\bf g}_P(s)} \cdot {\bf g}_P(\tau)\,  d s  d
\tau}\end{equation}
which, changing the integration limit, becomes 
  \begin{equation}\label{23new}
\displaystyle{\!\!\!\! \widetilde W \{ {\bf 0};  {\bf g}_P\}=
{1\over 2}\!
\int_{0}^{\infty}\!\!\!\! \int_{0}^{\infty}\!\!  {
 k(\vert{\tau} - s\vert) {\bf g}_P(s)} \cdot {\bf g}_P(\tau)
\, d s d \tau}~~.\end{equation} {
Hence, the definition of {\it finite thermal work process} follows as 
the set which comprises all those processes which correspond to 
a finite $\widetilde W\{ {\bf 0}; {\bf g}_P\}$, that is $\widetilde W\{ {\bf 0}; {\bf g}_P\}<\infty$.

According to \cite{SAM}, when the generic
case, characterised by any initial state
$\sigma(t)= (\theta(t),{\bf g}^t)$  prolonged via any process $P$, 
the thermal work can be expressed as follows
\begin{equation}\label{23h}
 \displaystyle{\widetilde W \{ {\bf g}^t;
{\bf g}_P\} \!=\! \int_{0}^{\infty}\!\!
\int_{0}^{\infty}\! \left[{ 1\over
2}{k(\vert{\tau} - s\vert) {\bf g}_P(s) } - 
k(\tau+s) {\bf g}^t(s) \right] \cdot {\bf g}_P(\tau)\, d s d \tau}~.
\end{equation}
\smallskip\noindent
The latter, on introduction of \footnote{In the definition \eqref{23i} the choice of the `-' 
minus sign is suggested by the consistency with the definition of $I$ given in \cite{SAM}. 
Indeed, definition  \eqref{23i} coincides with , (3.12) in \cite{SAM} when
the regular case, namely $\dot k\in L^1$, is considered.}
\begin{equation}\label{23i} \displaystyle{{\bf I}(\tau,{\bf g}^t):=
-  {\int_{0}^{\infty} {  k(\tau+s)\, {\bf g}^t (s) d
s}}
 }~~,
\end{equation}
gives
\begin{equation}\label{Wgen}
 \displaystyle{\widetilde W \{{\bf g}^t;  {\bf
g}_P\} \!=\! \int_{0}^{\infty} \left[ { 1\over 2}
\int_{0}^{\infty}\!\!\!\! {k(\vert{\tau} -
s\vert) {\bf g}_P(s)  d s} - {\bf
I}(\tau,{\bf g}^t) \right] \cdot {\bf
g}_P(\tau)d \tau}~~.
\end{equation}
Now, the setting is exactly the same in both the regular as well as in
the singular case, hence the result in \cite{SAM} can be repeated 
in an analogous way.  The difference consists in the definition 
\eqref{history} which, in the singular case, is given in terms of ${\bf g}^t$
while in the regular case (see (2.6) in \cite{SAM}) is expressed in terms of
 $ \bar{\bf g}^t$.

Following \cite{SAM}, on introduction of  Fourier transforms,
the {\it{ set  of all finite thermal work states}} is represented by the functional space
associated to finite  thermal work 
\begin{equation}\label{10}
\displaystyle{{\cal H}(\R^+,\R^3):=
\left\{ \phi:\R^+\!\to\!\R^3
 : \left\vert\int_{-\infty}^{+\infty}\!\! {
k_c(\omega) {\phi}_{+}(\omega) \cdot
   \overline{{\phi}_{+}(\omega)}   d
\omega} \right\vert < \infty\right\}}, \end{equation}
where $k_c(\omega)$ denotes  the Fourier cosine
transform\footnote{See for instance \cite{Ablo}.} 
of $k(\xi)\in L^1$, an even function of its real non negative
argument, $\xi=\vert{\tau} - s\vert$. 
The functional space
${\cal H}(\R^+,\R^3)$, can be equipped by the following
inner product
\begin{equation}\label{inn_Prod_k}
\displaystyle{\left<{ f}, \phi\right>_k:=
\int_{-\infty}^{+\infty} { k_c(\omega)\, { f}_{+}(\omega)
\cdot \overline{{\phi}_{+}(\omega)} ~ d
\omega}}
\end{equation}
and by the induced {\it { norm} }
\begin{equation}\label{norm} \displaystyle{\left\Vert \phi
\right\Vert_k := \left<\phi, \phi \right>_k=
\int_{-\infty}^{+\infty} { k_c(\omega)\, {\phi}_{+}(\omega) \cdot
   \overline{{\phi}_{+}(\omega)} ~  d
\omega}~~.}
\end{equation}
    On introduction of  the Fourier transforms\footnote{Some details are given in 
    Section 2 in  \cite{SAM}; background notions on Fourier transform are comprised, 
    for instance, in \cite{Ablo}.}
     of     ${\bf I}(\tau,{\bf g}^t)$ and of
    ${\bf g}_P$ defined via
    \begin{equation}\label{hat_I} 
        {\bf I}_{+}(\omega):=\!\!
    \displaystyle{\int_{0}^{\infty}\!\!\!
     {{\bf I}(\tau,{\bf g}^t) \,e^{-i\omega\tau}  d \tau }}~~,~~~
    {\bf g}_{+}(\omega):=\!\! \displaystyle{\int_{0}^{\infty}\!\!\!
     {{\bf g}_P(\tau) \,e^{-i\omega\tau}  d \tau }}~,
    \end{equation}
    Plancharel's theorem allows to evaluate the
    functional $\widetilde W$  in the dual 
    space under Fourier transform, as
    \begin{equation}\label{hat_work}
     \displaystyle{\!\!\widehat W \{{\bf g}^t;
    {\bf g}_P\}=\! {1 \over {2 \pi}}\left[
    \int_{-\infty}^{+\infty} \!\!\!\! k_c(\omega) {\bf g}_{+}(\omega)
    \cdot
       \overline{{\bf g}_{+}(\omega)}  \, d \omega - 
       \!\!\!\int_{-\infty}^{+\infty} \!\!\!\!
  {  {\bf I}_{+}(\omega) \cdot \overline{{\bf g}_{+}(\omega)} \, d \omega}\right]}
    \end{equation}
    where $\overline{{\bf g}_{+}(\omega)}$ denotes complex
    conjugate of the Fourier Transform ${\bf g}_{+}(\omega)$ of
    ${\bf g}_P(\tau)$. 
    As a special case,  $\widehat W\{{\bf 0}; {\bf g}_P\}$,  is given by 
    \begin{equation}\label{zero_hat_work}
     \displaystyle{\widehat W \{{\bf 0}; {\bf g}_P\}=\!
      {1 \over {2 \pi}} \int_{-\infty}^{+\infty}
    \!\!\!k_c(\omega)\, {\bf g}_{+}(\omega) \cdot
       \overline{{\bf g}_{+}(\omega)}  \, d
    \omega },
    \end{equation}
    which justifies the introduction \cite{SAM} of the space 
    ${\cal H}(\R^+,\R^3)$ in \eqref{10}. Furthermore, when
    \eqref{zero_hat_work} and \eqref{hat_work} are compared,
    it follows that, provided the first one is bounded, then the latter is 
    also bounded whenever
    \begin{equation}\label{2hat_I}
     \displaystyle{\left< 
     {\bf I}, {\bf g}\right>:=
    {1 \over {2 \pi}} \int_{-\infty}^{+\infty}
    {\bf I}_{+}(\omega) \cdot \overline{{\bf
    g}_{+}(\omega)} \, d \omega}~,
    \end{equation}
    where ${\bf I}_{+}(\omega)$ is given in (\ref{hat_I}), is finite. Hence, also in the singular case, the completion  with respect to the norm
$\Vert\cdot\Vert_k$ of the space ${\cal H}(\R^+,\R^3)$, denoted as 
${\cal H}_k(\R^+,\R^3)$, represents the set of all admissible thermal states 
according to the following definition \cite{SAM}.

 {\definition{
\qquad\qquad\qquad\qquad~{\bf admissible thermal states}

\noindent
The set of all
 admissible thermal states is
 the set of all states $\sigma(t)\in \Sigma_0$,
 such that, for any choice of the prolongation process $P$,
such that
${\bf g}_P\in{\cal H}_k$, it follows that ${\bf I}_{+}(\omega)$ belongs to the space 
\begin{equation}\label{H_prime}
\displaystyle{\!\!\!\!{\cal H}^{\prime}_k(\R^+,\R^3):=
\left\{
{f}:\R^+\!\to\!\R^3~
 {\rm  s.t.} ~\left\vert\left<{ f},
\phi\right>_k\right\vert < \infty, ~ \forall~\phi\in {\cal
H}_k(\R^+,\R^3) \right\}}~,
\end{equation}
dual of ${\cal H}_k(\R^+,\R^3)$ with respect  to
the inner product $\left<\cdot, \cdot\right>_k$
defined in (\ref{inn_Prod_k}).}}
\noindent
According to all the definitions here extended to the case of a material 
characterised by a singular heat flux relaxation function, the equivalence notion 
given in \cite{SAM} can be naturally extended to this kind of materials. 
This notion allows to single out all those thermal histories which are associated to the same heat flux. Notably,  according to Prop. 3.6 in \cite{SAM}, which 
holds true also in the singular case, if the heat flux which corresponds to  two different 
histories is the same, then also the corresponding thermal work is the same.
Specifically, given two arbitrary states $(\theta_1(t),{\bf g}^t_1)$ and
$(\theta_2(t),{\bf g}^t_2)$,   if and only if they 
$\forall~ {\bf g}_P~: [0, \tau) \to \R^3,~\forall~ \tau > 0$
the corresponding heat flux is the same, then also the thermal work is the same,
namely\footnote{The proof is exactly the same when both a regular as well as a
singular heat flux relaxation function are considered, hence, it is here omitted referring 
to Prop. 3.6 in \cite{SAM} and its proof therein.}
\begin{equation}\label{w-eq-eq}
\displaystyle{
 \widetilde Q \left\{\left({\bf g}_P \star{\bf
g}_1\right)^{t+\tau} \right\}\!=\! \widetilde Q
\left\{\left({\bf g}_P \star{\bf
g}_2\right)^{t+\tau} \right\}\Longleftrightarrow
\widetilde W \{ {\bf g}_1^t; {\bf g}_P\}
\!=\!\widetilde W \{ {\bf g}_2^t; {\bf
g}_P\}.}
\end{equation}
Hence, the state function
\begin{equation}\label{w-state}
\displaystyle{ {\cal W}(\sigma(t), {\bf g}_P): = \widetilde W \{
{\bf g}^t; {\bf g}_P\}
= \widehat W \{ {\bf g}^t; {\bf g}_P\}}~~,
\end{equation}
which represents the thermal work, can be introduced to stress that, under the 
physical viewpoint, it is the quantity of interest. Indeed, given a process $P$,
assigned via ${\bf g}_P$, the thermal work which is
associated to all those states identified by the same state function $\sigma(t)$,
is the same.
 Thus, if $\sigma
\in \Sigma$, it follows that both the heat flux
${\bf q}$, as well as the internal energy $e$,
given, in turn, by (\ref{2}) and (\ref{1}),
are finite.  According to the observation in \cite{FGR}, if a
conductor in a state $\sigma(0) \in \Sigma$ is
considered, its time evolution $\sigma(t)$ does
not necessarily belong to $\Sigma$ itself.
However, the existence of any thermodynamic
potential can be guaranteed only in the case when
${{\sigma(t) \in \Sigma}}$,  $\forall~ 0\le t<T$,
where $T$ denotes the duration of the process.
Hence, the study is restricted to those processes
which are physically meaningful, namely  admit an
associated finite thermodynamic potential. 

\section{Conclusion and perspectives} 
\label{section3}
\setcounter{equation}{0}
The present revisitation of the model of a rigid heat conductor with memory aims 
to provide suitable definitions to adapt the usual ones to the case when a less regular 
heat flux relaxation function is considered. Accordingly,   
the crucial physically  meaningful quantities are defined in a {\it slightly} modified 
manner to take into account the relaxed regularity requirements which are considered
in the case of a singular kernel heat flux relaxation function. 
The interesting conclusion is that most of the properties studied in \cite{SAM} remain valid as soon a suitable definition is adopted.  Notably, all the definitions 
which refer to a  heat flux relaxation which exhibits a singularity at $t=0$, coincide with the usual ones when the  heat flux relaxation is regular. Note that the minimisation procedure is not mentioned in the present article since the results  in \cite{SAM} are still valid: the only 
difference is that the thermal work and heat flux are expressed in the forms given in Sections 4 and 3, respectively. In addition, the minimisation  procedure  in \cite{SAM} conserves its validity and, hence, there is no need to repeat it. 
The importance of this generalisation relies in the perspectives it opens as far as
evolution problems are concerned. Indeed,  singular evolution problems are studied in
rigid thermodynamics  \cite{CVV2013a, [95]}, and also in isothermal viscoelasticity    \cite{CCVV, [96], CVV2013b}. Furthermore, inspired by new materials, regular as well as singular kernel problems in which the coupling a magnetic field with a viscoelastic behaviour is taken into account are investigated in \cite{CVV1, CVV2, CCVV, [94]}.
The results here presented may suggest a possible way to investigate the asymptotic behaviour 
of the temperature evolution in the case of a singular 
question still open.  

\begin{center} \bf Acknowledgments\end{center}
The financial  support of GNFM-INDAM, INFN and SAPIENZA Universit\`a di Roma, are gratefully acknowledged.

\end{document}